\documentclass[10pt,journal,cspaper,compsoc,onecolumn]{IEEEtran}
\usepackage{epsfig,endnotes}
\usepackage{amssymb}
\usepackage{graphicx}
\usepackage{amsfonts}
\usepackage{dsfont}
\usepackage{amsmath}
\usepackage{color}
\usepackage{booktabs}
\usepackage{epstopdf}
\usepackage{url}
\usepackage{tabularx}
\usepackage{verbatim} 
\usepackage{setspace}
\usepackage{caption}

\newtheorem{definition}{Definition}
\begin{document}

\title{On the Security of Privacy-Preserving Vehicular Communication Authentication with Hierarchical Aggregation and Fast Response}

\author{Lei Zhang,~\IEEEmembership{Member,~IEEE,} Chuanyan Hu, Qianhong Wu,~\IEEEmembership{Member,~IEEE,}
        Josep Domingo-Ferrer,~\IEEEmembership{Fellow,~IEEE,}
         and Bo Qin
\IEEEcompsocitemizethanks{\IEEEcompsocthanksitem Lei Zhang and Chuanyan Hu are with Shanghai Key Laboratory of Trustworthy Computing, Software Engineering Institute, East China Normal University, China; Qianhong Wu is with School of Electronic and Information Engineering, Beihang University, China; Josep Domingo-Ferrer is with the Department of Computer Engineering and Mathematics, Universitat Rovira i
Virgili, Catalonia; Bo Qin is with School of Information, Renmin University, China (e-mail: leizhang@sei.ecnu.edu.cn, chuanyanhu@ecnu.cn,
qhwu@xidian.edu.cn, josep.domingo@urv.cat, bo.qin@ruc.edu.cn).
}
\thanks{}}

\IEEEcompsoctitleabstractindextext{%
\begin{abstract}
In~\cite{IEEE-T-TC}, the authors proposed a highly efficient secure and
privacy-preserving scheme for secure vehicular communications. The proposed scheme consists of four protocols: system setup, protocol for
STP and STK distribution, protocol for common string
synchronization, and protocol for vehicular communications. Here
we define the security models for the protocol for
STP and STK distribution, and the protocol for vehicular communications,
respectively. We then prove that these two protocols are secure in our models.

\end{abstract}

}

\maketitle

\IEEEdisplaynotcompsoctitleabstractindextext

\IEEEpeerreviewmaketitle

\section{Security Model}

\subsection{Security Model for the Protocol for $STP$ and $STK$ Distribution}\label{model}

The security and privacy of the protocol for $STP$ and $STK$ distribution is defined in the game below. It is run between a challenger $CH$ and an adversary $Att$ who has full control of the network communications. $Att$ can be of
three types:
\begin{itemize}
  \item Type 1 adversary aims to break the message confidentiality property of our protocol. In our protocol, since we assume that the
underlying symmetric encryption/decryption scheme is secure, a type 1 adversary refers to an adversary who can violate the message confidentiality property of the
underlying signcryption scheme.
  \item Type 2 adversary aims to break the message authentication and non-repudiation properties of our protocol.
  \item Type 3 aims to break the privacy property of our protocol. Similar to a type 1 adversary, a type 3 adversary refers to an adversary who can violate the privacy property of the underlying signcryption scheme.
\end{itemize}

The game has the following stages:

\textbf{Initialize:} On input a security parameter $\ell$, $CH$ generates the system parameters $pub$ and passes $pub$ to $Att$.

\textbf{Attack:} According to the protocol for $STP$ and $STK$ distribution, at this stage, $Att$ is allowed to obtain the following information
from $CH$.
\begin{itemize}
  \item $Q_1$: The signcrypted message in the \textbf{Request} phase.
  \item $Q_2$: The de-signcrypted message in the \textbf{Verify} phase (in the case that an RSU is corrupted).
  \item $Q_3$: The ciphertext sent to the vehicle and the corresponding plaintext in the \textbf{Replay} and \textbf{Update} phases, respectively.
  \item	$Q_4$: For an identity-based system, usually we also allow $Att$ to obtain the (long-term) private keys of the vehicles and RSUs (except the target one(s)).
\end{itemize}

\textbf{Response:} This phase has three cases:
\begin{itemize}
  \item If $Att$ is of type 1, $Att$ returns two messages $(m_0, m_1)$ and an RSU's identity. $CH$ randomly chooses $m_b\in \{m_0, m_1\}$ and generates a signcrypted message $C$. We note that in our protocol, the vehicle's long-term pseudonym is included in the message. In $m_0$ and $m_1$, the vehicle's long-term pseudonyms are the same. $Att$ may continually make the queries in the \textbf{Attack} stage. $Att$ wins the game if he can distinguish
whether $C$ corresponds to $m_0$ or $m_1$ without querying the private key of the RSU or the plaintext corresponding to $C$.

  \item If $Att$ is of type 2, $Att$ returns a signcrypted message $C$
and an RSU's identity $ID_R$. Let $m=(n,LTP,\tau)$ be the plaintext
corresponding to $C$. $Att$ wins the game if $C$ can pass the Verify phase and $Att$ has never queried the private key corresponding to $LTP$ or the signcrypted message corresponding to $(m,LTP,ID_R)$.

  \item If $Att$ is of type 3, $Att$ returns two messages $m_0$ and $m_1$ and an RSU's identity. We note that in our protocol, the vehicle's long-term pseudonym is included in the message. Let the vehicles' long-term pseudonyms in $m_0$ and $m_1$ be $LTP_0$ and $LTP_1$, respectively. The only difference between $m_0$ and $m_1$ is that the two vehicles' long-term pseudonyms in $m_0$ and $m_1$ are different. $CH$ randomly chooses $m_b\in \{m_0, m_1\}$ and generates a signcrypted message $C$. $Att$ may continually make the queries in the \textbf{Attack} stage. $Att$ wins the game if he can distinguish whether $C$ corresponds to $LTP_0$ or $LTP_1$ without querying the private key of the RSU or
the plaintext corresponding to $C$.
\end{itemize}

\begin{definition}
The protocol for $STP$ and $STK$ satisfies message confidentiality if no type 1 adversary can win the above game in polynomial time with non-negligible probability.
\end{definition}

\begin{definition}
The protocol for $STP$ and $STK$ satisfies message authentication and non-repudiation if no type 2 adversary can win the above game in polynomial time with non-negligible probability.
\end{definition}

\begin{definition}
The protocol for $STP$ and $STK$ satisfies privacy if no type 3 adversary can win the above game in polynomial time with non-negligible probability.
\end{definition}

We note that the definition of privacy in this paper is slightly weaker than the definition of ciphertext anonymity (a stronger definition of privacy) in \cite{chen}. However, in our protocol, we do not need to consider the privacy of an RSU. Hence, our definition of privacy is sufficient for our protocol. Further, it is easy to see that if the protocol for $STP$ and $STK$ distribution achieves message confidentiality, then the protocol also achieves privacy.

\subsection{Security Model for Protocol for Vehicular Communications}

\indent The security of our protocol is modeled via the following
game between a challenger $CH$
and an adversary $Att$.

\textbf{Initialize:} On input a security parameter $\ell$, $CH$ generates the system parameters $pub$ and passes $pub$ to $Att$.

\smallskip \textbf{Attack}:  According to the protocol for vehicular communications, at this stage, $Att$ is allowed to obtain the following information from
$CH$.
\begin{itemize}
  \item $Q_5$: The short-term
    private key of a vehicle (corresponding to an identity-based system).

  \item $Q_6$: The signatures generated by the vehicles in the
\textbf{Sign} phase.

  \item $Q_7$: The real identity corresponding to a vehicle's
short-term pseudonym in the \textbf{Trace} phase.
\end{itemize}

We note that we do not need to model the signature verification and
aggregation procedures in the \textbf{Verify}, \textbf{Store}
and \textbf{Re-aggregate} phases, because $Att$ can do these operations
himself.

\medskip
 \textbf{Response}: In our protocol, since we assume that the
underlying symmetric encryption/decryption scheme is secure and the KGC is fully trusted, $Att$ cannot violate the privacy of a vehicle. Hence, $Att$ can
break our protocol if and only if he can output a forged
aggregate signature. Assume $Att$ outputs a set of $n$
vehicles' short-term
    pseudonyms from the set $L_{ID}^*=\{STP_1^*,...,STP_n^*\}$, $n$ messages
from the set
$L_m^*=\{m_1^*,...,m_n^*\}$, and an
aggregate signature $\sigma^*$.
We say that $Att$ wins the game if and only if
\begin{enumerate}
  \item $\sigma^*$ is a valid aggregate signature on messages $\{m_1^*,...,m_n^*\}$
     under $\{STP_1^*,...,STP_n^*\}$.

  \item At least one of the identities, without loss of generality, say $STP_1^*\in L_{ID}^*$ has not been
  submitted in the $Q_5$
  queries, and $(m_1^*, STP_1^*)$ has never been submitted
in the $Q_6$ queries.

\end{enumerate}

The above model captures the individual authentication and
non-repudiation properties of our protocol. As to the vehicle
privacy and traceability properties, they are achieved using short-term pseudonyms. This method is widely used in VANET systems.

\section{Security Proofs}
The security of our protocols is related to the bilinear Diffie-Hellman (BDH) and the computational Diffie-Hellman (CDH) problems.

Let $\mathbb{G}_1$, $\mathbb{G}_2$
be two additive cyclic groups and
$\mathbb{G}_T$ be a multiplicative cyclic group,
all with the same prime order $q$; $P_1$,
$P_2$ be random elements in $\mathbb{G}_1$ and $\mathbb{G}_2$,
respectively; $\psi$ be a computable
isomorphism from $\mathbb{G}_2$ to $\mathbb{G}_1$. A map $\hat e: \mathbb{G}_1 \times \mathbb{G}_2
\rightarrow \mathbb{G}_T$ is called bilinear if
1) $\hat{e}(a P_1,b P_2)=\hat{e}(P_1,P_2)^{ab}$ for any $a,b\in \mathbb{Z}/q\mathbb{Z}$;
2) $\hat{e}(P_1,P_2)\neq 1_{\mathbb{G}_T}$;
3) There exists an efficient
algorithm to compute $\hat{e}(P_1,P_2)$.

\begin{itemize}
  \item BDH$_{2,2,1}^\psi$ problem \cite{smart}: Given $(P_1,P_2,aP_2,bP_2,cP_1)$, compute $\hat{e}(P_1,P_2)^{abc}$ for unknown $a,b,c\in \mathbb{Z}/q\mathbb{Z}$.
  \item CDH$_{2,2,1}^\psi$ problem \cite{smart}: Given $(P_1,P_2,aP_2,bP_2)$,
 compute $abP_1$ for unknown $a,b\in \mathbb{Z}/q\mathbb{Z}$.
\end{itemize}


\subsection{Security of the Protocol for STP and STK Distribution}
Our results are all in the random oracle model. In each of the results below we assume that the adversary makes $q_{i}$~queries to $H_{i}$~for $i\in \{1,2,3,5\}$. Assume the numbers of $Q_1$ and $Q_2$ queries made by the adversary are denoted by $q_{s}$~and $q_{d}$, respectively.

\textbf{Theorem 1.} If a type 1 adversary wins the game defined in Section~\ref{model} with probability $\epsilon$, then a $CH$ running in polynomial time solves the BDH$_{2,2,1}^\psi$ problem with probability at least
\begin{displaymath}
\epsilon\cdot \frac{1}{q_{2}q_5}.
\end{displaymath}

\emph{Proof.}
Let $(P_1,P_2,aP_2,bP_2,cP_1)$ be the instance of the BDH$_{2,2,1}^\psi$ problem that we wish to solve.

\textbf{Initialize}: On input a security parameter $\ell$, $CH$ chooses $pub=(\mathbb{G}_1,\mathbb{G}_2,\mathbb{G}_T,\hat{e},P_1,P_2,U_1,U_2,\psi,H_1\sim
  H_6,$ $E_{k}(\cdot)/D_{k}(\cdot),$ $l_1,l_2,l_3,ID_{kgc},P_{kgc})$
      as the system public parameters, where $U_2=bP_2,U_1=\psi(U_2)$. We describe how $CH$ uses $Att$ to compute $\hat{e}(P_1,P_2)^{abc}$.

\medskip
\textbf{Attack}: $CH$ answers $Att$'s query as follows:

\medskip
$H_{1}(LTP_i)$ queries:\\
Choose $x_i$ at random from $\mathbb{Z}/q\mathbb{Z}$ and $k_i$ from the key space of $E_{k}(\cdot)/D_{k}(\cdot)$; compute $P_{V_i}= x_iP_1$; compute $LTK_i= xU_1$; store $(LTP_i,P_{V_i},LTK_{i},x_i,k_i)$~in $L_{1}$~and respond with $P_{V_i}$.

\medskip
$H_{2}(ID_{R_i})$ queries:\\
At the beginning of the simulation, choose $I$ uniformly at random from $\{1,...,q_2\}$.
We show how to respond to the $i$-th query made by $Att$ below. Note that we assume $Att$ does not make repeated queries.
\begin{itemize}
  \item If $i = I$~then respond with $aP_2$.
  \item Else choose $x_i'$ uniformly at random from $\mathbb{Z}/q\mathbb{Z}$; compute $P_{R_i}= x_i'P_2$; compute $B_i= x_i'U_2$; store $(ID_{R_i},P_{R_i},B_{i},x_i')$~in $L_{2}$~and respond with $P_{R_i}$.
\end{itemize}

\medskip
$H_{3}(Y_i||m_i)$ queries:
\begin{itemize}
  \item If $(Y_i,m_i,h_i)\in L_3$ for some $h_{i}$, return  $h_{i}$.
  \item Else choose $h_{i}$ uniformly at random from $\mathbb{Z}/q\mathbb{Z}$; add $(Y_i,m_i,h_i)$ to $L_{3}$ and return $h_{i}$.
\end{itemize}

\medskip
$H_{5}(\omega_i)$ queries:
\begin{itemize}
  \item If $(\omega_i,h_i') \in L_{5}$ for some $h_i'$, return $h_{i}'$.
  \item Else choose $h_i'$ uniformly at random from $\{0,1\}^{l_2}$; add $(\omega_i,h_{i}')$ to $L_{5}$ and return $h_{i}'$.
\end{itemize}

\medskip
$Q_4$ queries:\\
The input of this query is a pseudonym/identity of a vehicle/RSU.
We will assume that $Att$ makes the query $H_{1}(LTP_i)$/$H_{2}(ID_{R_i})$~before he makes the $Q_4$ query corresponding to $LTP_i$/$ID_{R_i}$.
\begin{itemize}
  \item If the input is equal to $ID_{R_I}$, abort the simulation.
  \item If the input is $LTP_i$, search $L_{1}$ for the entry $(LTP_i,P_{V_i},LTK_{i},x_i,k_i)$ and return $LTK_i$.
  \item Else search $L_{2}$~ for the entry $(ID_{R_i},P_{R_i},B_{i},x_i')$~corresponding to $ID_{R_i}$~and return $B_{i}$.
\end{itemize}

\medskip
$Q_1$ queries:\\
The input of this query is $(m_i,ID_{R_i})$, where $LTP_{i}$ is included in $m_i$.
We will assume $Att$ makes the queries $H_{1}(LTP_{i})$~and $H_{2}(ID_{R_i})$ before he makes this query.
\begin{itemize}
  \item Find the entry $(LTP_i,P_{V_i},LTK_{i},x_i,k_i)$ in $L_{1}$.
  \item Choose $r_i$ uniformly at random from $\mathbb{Z}_{q}^{*}$ and compute $Y_i= r_iP_{V_i}$.
  \item Compute $h_{i}= H_{3}(Y_i||m_i)$ (where $H_{3}$ is the simulator above).
  \item Compute $Z_i=(r_i + h_{i})LTK_{i}$.
  \item Compute $P_{R_i}= H_{2}(ID_{R_i})$ (where $H_{2}$ is the simulator above).
  \item Compute $\omega_i= \hat{e}(r_iLTK_{i},P_{R_i})$.
  \item Compute $y_i= H_{5}(\omega_i)\oplus (Z_i||m_i)$~(where $H_{5}$ is the simulator above).
  \item Return a signcrypted message $(Y_i,y_i)$.
\end{itemize}

\medskip
$Q_2$ queries:\\
The input of this query is a signcrypted message $(Y_i,y_i)$ and an identity of an RSU $ID_{R_i}$. We assume that $Att$ makes the query $H_{2}(ID_{R_i})$~ before making a $Q_2$ query. We have the following cases.

Case 1: $ID_{R_i} \neq ID_{R_I}$
\begin{itemize}
  \item Find the entry $(ID_{R_i},P_{R_i},B_{i},x_i')$~ in $L_{2}$.
  \item Compute $\omega_i = \hat{e}(Y_i,B_{i})$.
  \item If $\omega_i \not\in L_{5}$, return $\perp$; else compute $Z_i||m_i = y_i\oplus H_{5}(\omega_i)$.
  \item Let the pseudonym in $m_i$ be $LTP_i$. If $LTP_I \not\in L_{1}$, return $\perp$. Else compute $P_{V_i} = H_{1}(LTP_i)$.
  \item If $(Y_i,m_i) \notin L_{3}$, return $\perp$. Else compute $h_{i}= H_{3}(Y_i||m_i).$
  \item If $\hat{e}(Z_i,P_2) \neq \hat{e}(Y_i + h_{i}P_{V_i},U_2)$, return $\perp$. Else return $m_i,(Y_i,Z_i)$.
\end{itemize}

Case 2: $ID_{R_i} = ID_{R_I}$
\begin{itemize}
  \item Step through the list $L_{5}$ with entries $(w_i,h_{i}')$~as follows.
  \begin{itemize}
    \item Compute $Z_i||m_i= y_i\oplus h_{i}'$.
    \item Let the pseudonym in $m_i$ be $LTP_i$. If $LTP_i \in L_{1}$, let $P_{V_i}= H_{1}(LTP_{i})$ and find $LTK_{i}$ in $L_{1}$, else move to the next element in $L_{5}$ and begin again.
    \item If $(Y_i,m_i) \in L_{3}$, let $h_{i}= H_{3}(Y_i||m_i)$, else move to the next element in $L_{5}$.
    \item Check that $\omega_i= \hat{e}(Z_i - h_{i}LTK_{i},aP_2)$ and if not, move on to the next element in $L_{2}$ and begin again.
    \item Check that $\hat{e}(Z_i,P_2) = \hat{e}(Y_i + h_{i}P_{V_i},U_2)$, if so return $m_i, (Y_i,Z_i)$, else move on to the next element in $L_{5}$.
  \end{itemize}
  \item If no message has been returned after stepping through $L_{5}$, return $\perp$.
\end{itemize}

\medskip
$Q_3$ queries:\\
Find the corresponding symmetric key $k_i$ in $L_{1}$. Output the corresponding ciphertext or plaintext using $k_i$.

\medskip
\textbf{Response:} $Att$ outputs two identities $LTP^*,ID_{R}^*$ and two messages ${m_{0},m_{1}}$. If ${ID_{R}^* \neq ID_{R_I}}$, $CH$ aborts. Otherwise it chooses $y^{*} \in \{0,1\}^{l_2}$~and sets $Y^{*} = cP_1 $. It returns the signcrypted message $\sigma^{*} =(Y^{*},y^{*})$~to $Att$. $Att$ may continually make the queries in the \textbf{Attack} stage with the restriction defined in the model. These queries are answered in the same way as those made by $Att$ in the above stage.
At the end of this phase, $Att$ outputs a bit $b$. $CH$ searches $L_{1}$~for the entry $(LTP^*,P_{V}^*,LTK^*,{x}^*,k^*)$, she chooses some $\omega^*$ at random from $L_{5}$ and returns
  \begin{displaymath}
  {\omega^*}^{{{x}^*}^{-1}}
  \end{displaymath}
  as her guess at the solution to the BDH$_{2,2,1}^\psi$ problem.

In the above simulation, if $CH$ does not abort, then $Att$'s view is identical to the real-world attack.
Similar to the security proof of Theorem 2 in \cite{chen}, we have that
$CH$ does not abort with
probability at least
        \begin{displaymath}
        \frac{1}{q_{2}}.
        \end{displaymath}
Since $\omega^*$ is randomly chosen from $L_{5}$,
we have that the possibility for $CH$ to solve the BDH$_{2,2,1}^\psi$ problem is at least
        \begin{displaymath}
        \epsilon\cdot  \frac{1}{q_{2}q_5}.
        \end{displaymath}

\textbf{Theorem 2.} If a type 2 adversary wins the game defined in Section~\ref{model} with probability $\epsilon$, then a $CH$ running in polynomial time solves the CDH$_{2,2,1}^\psi$ problem with probability at least
\begin{displaymath}
\epsilon^2(1 - \frac{q_{s}(q_{3} + q_{s})}{q})^2\cdot \frac{1}{4q_{1}^2(q_{3} + q_{s})^2}.
\end{displaymath}

\emph{Proof.} Let $(P_1,P_2,aP_2,bP_2)$ be the instance of the CDH$_{2,2,1}^\psi$ problem that we wish to solve.

\textbf{Initialize}: On input a security parameter $\ell$, $CH$ chooses $pub=(\mathbb{G}_1,\mathbb{G}_2,\mathbb{G}_T,\hat{e},P_1,P_2,U_1,U_2,\psi,H_1\sim
H_6,$ $E_{k}(\cdot)/D_{k}(\cdot),$ $l_1,l_2,l_3,ID_{kgc},P_{kgc})$
as the system public parameters, where $U_2=bP_2,U_1=\psi(U_2)$. We describe how $CH$ uses $Att$ to compute $abP_1$.

\medskip
\textbf{Attack}: $CH$ answers $Att$'s query as follows:

\medskip
 $H_{1}(LTP_{i})$ queries:\\
 At the beginning of the simulation, choose $I$~uniformly at random from $\{1,...,q_{1}\}$. Note that we assume $Att$ does not make
repeated queries.
 \begin{itemize}
   \item If $i = I$ then respond with $H_{1}(LTP_{i})= \psi(aP_2)$; choose $k_i$ from the key space of $E_{k}(\cdot)/D_{k}(\cdot)$; store $(LTP_i,P_{V_i},\bot,$ $\bot,k_i)$~in $L_{1}$.
   \item Else choose $x_i$ uniformly at random from $\mathbb{Z}/q\mathbb{Z}$ and $k_i$ from the key space of $E_{k}(\cdot)/D_{k}(\cdot)$; compute $P_{V_i}= x_iP_1$; compute $LTK_{i}= x_iU_{1}$; store $(LTP_{i},P_{V_i},LTK_{i},x_i,k_i)$ in $L_{1}$ and respond with $P_{V_i}$.
 \end{itemize}

\medskip
 $H_{2}(ID_{R_i})$ queries:\\
Choose $x_i'$ uniformly at random from $\mathbb{Z}/q\mathbb{Z}$;
compute $P_{R_i}= x_i'P_2$; compute $B_i= x_i'U_2$; store $(ID_{R_i},P_{R_i},B_{i},x_i')$~in $L_{2}$~and respond with $P_{R_i}$.

\medskip
$H_{3}(Y_i||m_i)$ queries:
\begin{itemize}
  \item If $(Y_i,m_i,h_i)\in L_3$ for some $h_{i}$, return $h_{i}$.
  \item Else choose $h_{i}$ uniformly at random from $\mathbb{Z}_{q}^{*}$; add $(Y_i,m_i,h_i)$ to $L_{3}$ and return $h_{i}$.
\end{itemize}

\medskip
$H_{5}(\omega_i)$ queries:
\begin{itemize}
  \item If $(\omega_i,h_i') \in L_{5}$ for some $h_i'$, return $h_{i}'$.
  \item Else choose $h_i'$ uniformly at random from $\{0,1\}^{l_2}$; add $(\omega_i,h_{i}')$ to $L_{5}$ and return $h_{i}'$.
\end{itemize}

\medskip
$Q_4$ queries:\\
The input of this query is a pseudonym/identity of a vehicle/RSU.
We will assume that $Att$ makes the query $H_{1}(LTP_i)$/$H_{2}(ID_{R_i})$~before he makes the $Q_4$ query corresponding to $LTP_i$/$ID_{R_i}$.
\begin{itemize}
  \item If the input is equal to $LTP_{I}$, abort the simulation.
  \item Else if the input is $LTP_i$, search $L_{1}$ for the entry $(LTP_i,P_{V_i},LTK_{i},x_i,k_i)$ and return $LTK_i$.
  \item Else search $L_{2}$ for the entry $(ID_{R_i},P_{R_i},B_{i},x_i')$~corresponding to $ID_{R_i}$~and return $B_{i}$.
\end{itemize}

\medskip
$Q_1$ queries:\\
The input of this query is $(m_i,ID_{R_i})$, where $LTP_{i}$ is included in $m_i$.
We will assume that $Att$ makes the queries $H_{1}(LTP_{i})$ and $H_{2}(ID_{R_i})$ before he makes this query. Two cases arise:

Case 1: $LTP_{i} \neq LTP_{I}$\\
Use the simulator of $Q_1$ in the proof of Theorem 1.

Case 2: $LTP_{i} = LTP_{I}$
\begin{itemize}
  \item Choose $r_i, h_i$ uniformly at random from $\mathbb{Z}_{q}^{*}$.
  \item Compute $Y_i= r_iP_1 - h_{i}H_1(LTP_{i})$ and $Z_i= r_iU_1$.
  \item Add $(Y_i,m_i,h_{i})$ to $L_{3}$.
  \item Find the entry $(ID_{R_i},P_{R_i},B_{i},x_i')$~ in $L_{2}$.
  \item Compute $\omega_i=\hat{e}(Y_i,B_{i})$.
  \item Compute $y_i= H_{5}(\omega_i)\oplus (Z_i||m_i)$~(where $H_{5}$ is the simulator above).
  \item Return $(Y_i,y_i)$.
\end{itemize}

\medskip
$Q_2$ queries:
\begin{itemize}
  \item Find the entry $(ID_{R_i},P_{R_i},B_{i},x_i')$~ in $L_{2}$.
  \item Compute $\omega_i = \hat{e}(Y_i,B_{i})$.
  \item If $\omega_i \not\in L_{5}$, return $\perp$; else find $y_i$ corresponding to $\omega_i$ and compute $Z_i||m_i = y_i\oplus H_{5}(\omega_i)$.
  \item Let the pseudonym in $m_i$ be $LTP_i$. If $LTP_I \not\in L_{1}$, return $\perp$. Else compute $P_{V_i} = H_{1}(LTP_i)$.
  \item If $(Y_i,m_i) \not\in L_{3}$, return $\perp$. Else compute $h_{i}= H_{3}(Y_i||m_i).$
  \item If $\hat{e}(Z_i,P_2) \neq \hat{e}(Y_i + h_{i}P_{V_i},U_2)$, return $\perp$. Else return $m_i,(Y_i,Z_i)$.
\end{itemize}

\medskip
$Q_3$ queries:\\
Find the corresponding symmetric key $k_i$ in $L_{1}$. Output the corresponding ciphertext or plaintext using $k_i$.

\medskip
In the above simulation, if $CH$ does not abort, then $Att$'s view is identical to the real-world attack.
Similar to the security proof of Theorem 3 in \cite{chen}, we have $CH$ does not abort with probability at least
\begin{displaymath}
 (1 - \frac{q_{s}(q_{3} + q_{s})}{q})\cdot \frac{1}{q_{1}}.
\end{displaymath}

With probability
\begin{displaymath}
 \epsilon(1 - \frac{q_{s}(q_{3} + q_{s})}{q})\cdot \frac{1}{q_{1}}
\end{displaymath}
$Att$ outputs a forgery $m^*,(Y^*,Z^*)$, where the pseudonym in $m^*$ is $LTP_I$.

\medskip
\textbf{Response:}
  According to the Splitting Lemma, $CH$ replays $Att$ with the same random tape but different choice of
the response of $H_3$. With probability
\begin{displaymath}
 \epsilon^2(1 - \frac{q_{s}(q_{3} + q_{s})}{q})^2\cdot \frac{1}{4q_{1}^2(q_{3} + q_{s})^2}
\end{displaymath}
the two runs yield two forgeries $m^*,(Y^*,Z^*)$ and $m^*,(Y^*,\hat{Z}^*)$ with $Z^*\neq \hat{Z}^*$ and $h^*\neq \hat{h}^*$, where $h^*$ and $\hat{h}^*$ are the outputs of $H_3$ corresponding to $(Y^*,m^*)$ in the first and second runs of the simulation respectively. Let $P_{V^*}=H_1(LTP_I)$.
Since the two forgeries should be valid, we have
$$\hat{e}(Z^*,P_2) = \hat{e}(Y^* + h^*P_{V^*},U_2)$$ and $$\hat{e}(\hat{Z}^*,P_2) = \hat{e}(Y^* + \hat{h}^*P_{V^*},U_2).$$
Since $P_{V^*}=aP_1$, we have $$abP_1={(h^*-\hat{h}^*)^{-1}}(Z^*-\hat{Z}^*).$$

\textbf{Theorem 3.} If a type 3 adversary wins the game defined in Section~\ref{model} with probability $\epsilon$, then a $CH$ running in polynomial time solves the BDH$_{2,2,1}^\psi$ problem with probability at least
\begin{displaymath}
\epsilon\cdot \frac{1}{q_{2}q_5}.
\end{displaymath}

\emph{Proof.} The proof is the same as that of the Theorem 1.

\subsection{Security of the Protocol for Vehicular Communications}
 In each of the results below we assume that the adversary makes $q_{H_i}$~queries to $H_{i}$ for $i\in \{1,2,3\}$. We assume $Att$ can ask at most
 $q_K$ times $Q_5$ queries, and $q_S$ times $Q_6$ queries.

\textbf{Theorem 4.}
If there exists an adversary
$Att$ who has an advantage $\varepsilon$ to break our protocol, then the CDH$_{2,2,1}^\psi$
problem can be solved in polynomial time
with probability at least
$$\varepsilon'\geq (1-\frac{1}{q_{H_1}})^{q_K}
(1-\frac{1}{q_{H_1}}\frac{1}{q_{H_2}}(1-\frac{1}{q_{H_3'}}))^{q_S}\frac{1}{q_{H_1}}\frac{1}{q_{H_2}}(1-\frac{1}{q_{H_3'}})
\varepsilon.$$

\noindent\emph{Proof.} Let $(P_1,P_2,aP_2,bP_2)$ be the instance of the CDH$_{2,2,1}^\psi$ problem that we wish to solve.

\smallskip
\textbf{Initialize}: On input a security parameter $\ell$, $CH$ chooses $pub=(\mathbb{G}_1,\mathbb{G}_2,\mathbb{G}_T,\hat{e},P_1,P_2,U_1,U_2,\psi,H_1\sim
H_6,$ $E_{k}(\cdot)/D_{k}(\cdot),$ $l_1,l_2,l_3,ID_{kgc},P_{kgc})$
as the system public parameters  and $\lambda$ from the key space of $E_{k}(\cdot)/D_{k}(\cdot)$, where $U_2=bP_2,U_1=\psi(U_2)$. We describe how $CH$ uses $Att$ to compute $abP_1$.

\smallskip
 \textbf{Attack:} $CH$ answers $Att$'s query as follows:

\medskip
$H_1(STP_i,j)$ queries:\\
Let $\bf H_1$ be the list of previous answers to these queries.
$CH$ picks $I\in [1, q_{H_1}]$
uniformly at random. Whenever $CH$ receives an $H_1$ query
on $(STP_i,j)$ for $j\in\{0,1\}$, $CH$ does the following:
\begin{enumerate}
  \item If there is a tuple
  $(STP_k,\alpha_{k,0},\alpha_{k,0}',\alpha_{k,1},\alpha_{k,1}',P_{k,0},P_{k,1})$
  on the list $\bf H_1$ such that $STP_i=STP_k$, return $P_{k,j}$ as the answer.

  \item Else if $i=I$, randomly choose
  $\alpha_{i,0},\alpha_{i,0}',\alpha_{i,1},\alpha_{i,1}'\in \mathbb{Z}/q\mathbb{Z}$, set
$P_{i,0}=\alpha_{i,0}P_1+\alpha_{i,0}'U_1,
P_{i,1}=\alpha_{i,1}P_1+\alpha_{i,1}'U_1$, add
$(STP_i,\alpha_{i,0},\alpha_{i,0}',\alpha_{i,1},\alpha_{i,1}',P_{i,0},P_{i,1})$
to $\bf H_1$ and return $P_{i,j}$ as the answer.

  \item Else set $\alpha_{i,0}'=0,\alpha_{i,1}'=0$,
randomly choose $\alpha_{i,0},\alpha_{i,1}\in \mathbb{Z}/q\mathbb{Z}$, set
$P_{i,0}=\alpha_{i,0}P_1,P_{i,1}=\alpha_{i,1}P_1$, add
$(STP_i,\alpha_{i,0},\alpha_{i,0}',$ $\alpha_{i,1},\alpha_{i,1}',P_{i,0},P_{i,1})$
to $\bf H_1$ and return $P_{i,j}$ as the answer.
\end{enumerate}

\medskip $H_2(CS_i)$ queries:\\
Let $\bf H_2$ be the list of previous answers to these queries.
$CH$
picks $J\in [1, q_{H_2}]$ uniformly at random. Whenever $Att$ issues a query $H_2(CS_i)$, the same answer from the list
$\bf H_2$ will be given if the request has been asked before.
Otherwise, $CH$ selects a random $\beta_i\in \mathbb{Z}/q\mathbb{Z}$; if
$i=J$, computes $\hat{P}_{CS_i}=\beta_iP_2$, else sets $\hat{P}_{CS_i}=\beta_iaP_2$. Finally,
$CH$ adds $(CS_i,\hat{P}_{CS_i},\beta_i)$ to $\bf H_2$ and
returns $\hat{P}_{CS_i}$ as the answer.

\medskip
$H_3(m_i,STP_i,CS_i)$ queries:\\
Let $\bf H_3$ be the list of previous answers to these queries.
 Whenever $Att$ issues a query
$(m_i,STP_i,CS_i)$ to $H_3$, the same answer from the list
$\bf H_3$ will be given if the request has been asked before.
Otherwise, $CH$ first submits $(STP_i,0)$ to $H_1$,
then finds the tuple
$(STP_i,\alpha_{i,0},\alpha_{i,0}',\alpha_{i,1},\alpha_{i,1}',P_{i,0},P_{i,1})$
on $\bf H_1$, and finally does the following:
\begin{enumerate}
  \item If $STP_i=STP_I$ and $CS_i=CS_J$ (we assume that $Att$
  can ask at most $q_{H_3'}<q_{H_3}$ times such kind of queries),
  randomly choose $K\in [1,q_{H_3'}]$.
 \begin{enumerate}
   \item If it is the $K$-th query,
   set $c_i=-\alpha_{i,0}'/\alpha_{i,1}'$, add $(m_i,STP_i,CS_i,c_i)$
        to $\bf H_3$ and return $c_i$.

   \item Else select a random $c_i\in \mathbb{Z}/q\mathbb{Z}$, add $(m_i,STP_i,CS_i,c_i)$
        to $\bf H_3$ and return $c_i$ as the answer.
 \end{enumerate}

  \item Else, select a random $c_i\in \mathbb{Z}/q\mathbb{Z}$, add $(m_i,STP_i,CS_i,c_i)$
        to $\bf H_3$ and return $c_i$ as the answer.
\end{enumerate}

\medskip
 $Q_5$ queries: When $Att$ issues a private key query corresponding to $STP_i$, the same answer will be given if
the request has been asked before. Otherwise, $CH$
looks for a tuple
$(STP_i,\alpha_{i,0},\alpha_{i,0}',\alpha_{i,1},\alpha_{i,1}',$
$P_{i,0},P_{i,1})$ on $\bf H_1$;
if none is found, $CH$ makes an
$H_1$ query on $(ID_i,j)$ ($j=0$ or 1) to generate such a tuple,
and finally does as follows
\begin{enumerate}
  \item If $ID_i=ID_I$, abort.
  \item Else return
    $(D_{i,0},D_{i,1})$ as the answer, where $D_{i,0}=\alpha_{i,0}U_1,D_{i,1}=\alpha_{i,1}U_1$.

\end{enumerate}

\medskip
$Q_6$ queries: The input of this query is
$(CS_i,m_i,STP_i)$; $CH$ first makes $H_1(ID_i,0),H_2(CS_i)$ and
$H_3(m_i,STP_i,CS_i)$ queries if they have not been made
before, then recovers
$(STP_i,\alpha_{i,0},\alpha_{i,0}',\alpha_{i,1},\alpha_{i,1}',P_{i,0},P_{i,1})$
from $\bf H_1$, $(CS_i,\hat{P}_{CS_i},\beta_i)$ from $\bf H_2$,
$(m_i,STP_i,CS_i,c_i)$
from $\bf H_3$ and
generates the signature as follows
\begin{enumerate}
    \item If $STP_i=STP_I,CS_i=CS_J$, and
    $c_i=-\alpha_{i,0}'/\alpha_{i,1}'$,
       choose $S_{i,2}\in G_1^*$, compute
    $S_{i,1}=\beta_iS_{i,2}+\alpha_{i,0}U_1+\alpha_{i,1}c_iU_1$, output $M_i=(m_i||STP_i||(S_{i,1},S_{i,2}))$.

    \item Else if $STP_i=STP_I,CS_i=CS_J$, abort.

    \item  Else if $STP_i=STP_I$, choose $r_i\in \mathbb{Z}/q\mathbb{Z}$,
    set $S_{i,2}=r_iP_1-\beta_i^{-1}(P_{i,0}+c_iP_{i,1})$, compute
    $S_{i,1}=r_i\psi(\hat{P}_{CS_1})$, output $M_i=(m_i||STP_i||(S_{i,1},S_{i,2}))$.

    \item Else, randomly choose $r_i\in \mathbb{Z}/q\mathbb{Z}$, compute $S_{i,2}=r_iP_1$, set
    $S_{i,1}=r_i\psi(\hat{P}_{CS_i})+\alpha_{i,0}U_1+c_i\alpha_{i,1}U_1$,
    output $M_i=(m_i||STP_i||(S_{i,1},S_{i,2}))$.
\end{enumerate}

Note that in the protocol, $CS_i$ is only for one-time use.
Hence, it is reasonable for $CH$ to abort when
$STP_i=STP_I,CS_i=CS_J$ and $c_i\neq
-\alpha_{i,0}'/\alpha_{i,1}'$.

\medskip
$Q_7$ queries: $CH$ outputs the real identity of a vehicle based on the \textbf{Trace} phase using $\lambda$.

\medskip
\noindent \textbf{Response:} Eventually, $Att$ returns
$L_{ID}^*=\{STP_1^*,...,STP_n^*\}$; $n$ messages from the set
$L_M^*=\{m_1^*,...,m_n^*\}$; a common string $CS^*$ and a
forged aggregate signature $\sigma^*=(S_{1}^*,S_{2}^*)$.

$CH$ recovers $(STP_i^*,\alpha_{i,0}^*,{\alpha_{i,0}'}^*,\alpha_{i,1}^*,{\alpha_{i,1}'}^*,P_{i,0}^*,P_{i,1}^*)$
from $\bf H_1$, $(CS^*,\hat{P}_{CS^*},\beta^*)$ from $\bf H_2$,
$(m_i^*,STP_i^*,CS^*,c_i^*)$
from $\bf H_3$
for all $i,1\leq i\leq n$.

$CH$ requires that $CS^*=CS_J$ and there exists $i\in
\{1,...,n\}$ such that $STP_i^*=STP_I$, $c_i^*\neq
-\alpha_{i,0}'^*/\alpha_{i,1}'^*$ and $Att$ has not made a $Q_6$ query on
$(CS^*,m_i^*,STP_i^*)$. Without loss of generality, we
let $i=1$. In addition, the forged aggregate signature must satisfy
$$\hat{e}(S_1^*,P_2)=\hat{e}(S_2^*,\hat{P}_{CS^*})\hat{e}(\sum_{i=1}^nP_{i,0}^*+\sum_{i=1}^n c_i^*
P_{i,1}^*,U_2).$$ Otherwise, $CH$ aborts.

If $CH$ does not abort,
by our setting, $P_{1,0}^*=\alpha_{1,0}^*P_1+\alpha_{1,0}'^*U_1$,
$P_{1,1}^*=\alpha_{1,1}^*P_1+\alpha_{1,1}'^*U_1$, $\hat{P}_{CS^*}=\beta^*P_2$; and for $i,2\leq i\leq n$,
$P_{i,j}^*=\alpha_{i,j}^*P_1$, where $j\in \{0,1\}$; hence, $CH$ can compute

$$abP_1=(\alpha_{1,0}'^*+ c_1^*\alpha_{1,1}'^*)^{-1}(S_1^*- \sum_{i=2}^n
\alpha_{i,0}^*U_1- \sum_{i=2}^n \alpha_{i,1}^*c_i^*
U_1-\beta^*S_2^*-(\alpha_{1,0}^*+ c_1^*\alpha_{1,1}^*)U_1).$$

To complete the proof, we shall show that $CH$ solves the
given instance of the CDH$_{2,2,1}^\psi$ problem with probability at least
$\varepsilon'$. First, we analyze the four events needed for
$CH$ to succeed:

\begin{itemize}
  \item $\Sigma$ 1: $CH$ does not abort in the above simulation.

  \item $\Sigma$ 2: $Att$ generates a valid and nontrivial aggregate signature
forgery.
  \item $\Sigma$ 3: Event $\Sigma$ 2 occurs, $CS^*=CS_J$ and there exists $i\in \{1,...,n\}$ such
that $STP_i^*=STP_I$, $c_i^*\neq -\alpha_{i,0}'^*/\alpha_{i,1}'^*$ (as
mentioned previously, we assume $i=1$).
\end{itemize}

$CH$ succeeds if all of these events happen. The
probability $\Pr[\Sigma\ 1\wedge \Sigma\ 2\wedge \Sigma\ 3]$ can be decomposed as
\begin{eqnarray*}
&&\Pr[\Sigma\ 1\wedge \Sigma\ 2\wedge \Sigma\ 3]\\
 &=&
\Pr[\Sigma\ 1]\Pr[\Sigma\ 2|\Sigma\ 1]\Pr[\Sigma\ 3|\Sigma\ 1 \wedge
\Sigma\ 2].
\end{eqnarray*}

\indent It is easy to see that the above probability for $CH$ to solve the CDH$_{2,2,1}^\psi$ problem is
\begin{eqnarray*}
\varepsilon'&=&\Pr[\Sigma\ 1\wedge \Sigma\ 2\wedge \Sigma\ 3] \\
&\geq& (1-\frac{1}{q_{H_1}})^{q_K}
(1-\frac{1}{q_{H_1}}\frac{1}{q_{H_2}}(1-\frac{1}{q_{H_3'}}))^{q_S}\frac{1}{q_{H_1}}\frac{1}{q_{H_2}}(1-\frac{1}{q_{H_3'}})
\varepsilon.
\end{eqnarray*}


\end{document}